\documentclass[10pt]{article}

\usepackage{amssymb}
\usepackage{graphicx}
\usepackage[numbers,compress]{natbib}
\begin{document}

	\title{New charged dynamical particles in spatially flat FLRW space-times}
	
	\author{Ion I. Cot\u aescu\thanks{Corresponding author E-mail:~~i.cotaescu@e-uvt.ro}\\
		{\it West University of Timi\c soara,} \\{\it V. Parvan Ave. 4,
			RO-300223 Timi\c soara}}
	
	\maketitle

\begin{abstract}
New  time-dependent metric tensors with spherical symmetry satisfying  the Einstein-Maxwell equations  in space-times with FLRW asymptotic behaviour are  derived here for the first time. These   geometries describe dynamical charged non-rotating black holes hosted by the perfect fluid of the asymptotic FLRW space-times. Their gravitational sources are the stress-energy tensors formed by a contribution of the perfect fluid and an electromagnetic one due to  the Coulomb field produced by the time-dependent black-hole charge in the asymptotic FLRW background.  The dynamics of these models is  determined by the dynamical mass, which may be an arbitrary function of time, and two arbitrary real-valued parameters.  The first one simulates the effect of a cosmological constant as in our $\kappa$-models we proposed recently   [I. I. Cotaescu, Eur. Phys. J. C (2024) 84:819]. The second parameter relates surprisingly the dynamical black hole charge to the cubic root of the mass function. The role of these parameters is investigated analyzing simple examples of dynamical charged black holes in the matter-dominated universe.

Pacs: 04.70.Bw
\end{abstract}


\section{Introduction}
In the actual cosmology, the  Friedmann- Lema\^ itre-Robertson-Walker (FLRW) space-times are the principal models of universe  that may be populated  either with static black holes, which are vacuum solutions of Einstein's equations \cite{BH}, or with dynamical particles defined as isotropic solutions of Einstein's equations with perfect fluid laying out a FLRW asymptotic behaviour (see for instance Ref. \cite{TB1}).  The first metric of dynamical particles behaving as a black holes was proposed by McVittie long time ago \cite{MV} and then studied by many authors \cite{MV1,MV2,MV3,MV4}.  These geometries lay out curved space sections being produced by the pressure of the perfect fluid which is singular on the Schwarzschild sphere while its density remains just that of the asymptotic FLRW space-time.

As an alternative,  we initiated recently the study of a new type of spherically symmetric dynamical particles which are exact solutions of Einstein's equations with perfect fluid having singular densities but preserving the fluid pressure of the asymptotic FLRW space-times that  this time have flat space sections  \cite{Cot}. We have shown that these geometries behave as non-rotating black holes in the physical space domain bordered by the black hole and cosmological dynamical horizons. Moreover, these dynamical particles give rise to photon spheres and black hole shadows as in the case of the static Schwarzschild black holes \cite{Cot}.  For this reason we said that these are models of  Schwarzschild-type dynamical particles or black holes. Subsequently we generalized these models obtaining new solutions of Einstein's equations with perfect fluid depending on a time-dependent  mass and a real valued parameter $\kappa$ giving more flexibility to these models, we called simply the $\kappa$-models \cite{Cot1}. Moreover, we have shown that the models with $\kappa\not=0$ may describe evaporating black holes in expanding universes \cite{Cot2}. The Schwarzschild-type models we proposed initially \cite{Cot} can be seen now as particular cases with $\kappa=0$. 

In the present paper we would like to extend our study looking for geometries of charged dynamical black holes in which the Einstein tensors are proportional with stress-energy tensors gathering the contribution of a perfect fluid and that of the Coulomb field produced by the dynamical black hole charge in the asymptotic FLRW space-time. In this manner we follow the same strategy as in the case of the genuine Reissner-Nordstr\" om black hole  whose Coulomb field is calculated in the asymptotic Minkowski space-time \cite{RN1,RN2,RN3,RN4}. We start with the metric tensors of our $\kappa$-models where we introduce Coulomb terms with time-dependent charges,  according to our previous results concerning the Maxwell field in FLRW space-times  \cite{CM}.  In this manner we obtain geometries with diagonal Einstein tensors but in which we cannot separate the Coulomb contribution from that of the perfect fluid. Surprisingly, this can be done if and only if  the dynamical charge is proportional to the cubic root of the mass function through a new parameter we denote by $\lambda$. Thus we have to speak about $(\kappa,\lambda)$-models that generalize the our previous $\kappa$-models of neutral black holes.   Our goal is to study here the special features of the models with $\lambda\not=0$ pointing out a natural limitation of the black hole charge and the specific new behaviour of the black hole horizon.

We start our study in the next section presenting first the physical frames with Painlev\' e-Gullstrand  coordinates  \cite{Pan,Gul}  which are suitable for studying the dynamical black holes in co-moving proper frames where the Einstein tensors are diagonal. We give the form of the Einstein-Maxwell equations in these frames using the Coulomb field derived in Ref. \cite{CM}.  Moreover, we identify the scale factors of the asymptotic FLRW space-times separating the contribution of their perfect fluids. Section 3 is devoted to our new dynamical solutions of the Einstein-Maxwell equations. These are constructed adding the specific Coulomb term in the metric tensors of our previous $\kappa$-models. These terms keep the diagonal form of the Einstein tensors but the separation of the electromagnetic stress-energy tensor can be done only if the dynamical charge is proportional to the cubic root of the mass function though the mentioned constant, $\lambda$. Another specific feature of these models is that the metric tensor can take complex values in a dynamical sphere, we call here the prohibited sphere, as this must be eliminated from the space-time domain. In the next section we study particular examples for which the radii of the cosmological and black hole horizons can be derived analytically, solving a simple cubic equation. We show why  the interval of the parameter $\lambda$   is finite, limiting  thus the charge carried by the black hole.  Finally, we find that, in the   models with $\lambda\not= 0$,   the black hole horizon, which arises simultaneously with the cosmological one on the same sphere,  disappears later  when the prohibited sphere takes over its role. In the last section we present our concluding remarks.

The method of solving cubic equations is given  in Appendix.  We use the Planck units with $\hbar=c=G=1$.

\section{Einstein-Maxwell equations in physical frames}

The static or dynamic non-rotating black holes with spherical symmetry are studied mainly in isotropic space-times with spatially flat sections. In these manifolds it is convenient to use the physical frames $\{t, {\bf x}\}$  with coordinates of  Painlev\' e-Gullstrand  type  \cite{Pan,Gul},  $x^{\mu}$ ($\alpha,\mu,\nu,...=0,1,2,3$). These are formed by the {\em cosmic time}, $x^0=t$, and physical Cartesian space coordinates,  ${\bf x}=(x^1,x^2,x^3)$, associated to the spherical ones $(r,\theta,\phi)$. For example, in the physical frame of the Reissner-Nordstr\" om-de Sitter black hole, the line element takes the form 
\begin{equation}
	ds^2=f(r)dt^2+2\sqrt{1-f(r)}\,dtdr-dr^2 -r^2d\Omega^2\,,\label{ss}
\end{equation}	
where 
\begin{equation}\label{frSdS}
	f(r)=1-\frac{2m}{r}+\frac{q^2}{r^2}-\omega_{H}^2 r^2\,,
\end{equation} 
depends on the black hole mass $m$, charge $q$ and Hubble de Sitter constant frequency $\omega_{H}$.  

Eq. (\ref{ss}) suggests the substitution $f(r)=1-h(r)^2$ helping us to simplify the off-diagonal components of the metric tensor. Therefore, in what follows  we consider that the line elements in physical frames of the dynamical manifolds have the general form  
\begin{eqnarray}
	ds^2&=& g_{\mu\nu}(x)dx^{\mu}dx^{\nu}=dt^2 -\left[dr-h (t,r)dt\right]^2-r^2 d\Omega^2 \nonumber\\
	&=&\left[1-h (t,r)^2\right]dt^2+2 h(t,r)dr dt -dr^2-r^2 d\Omega^2\,,\label{fam}
\end{eqnarray} 
depending on the smooth functions $h$ and $d\Omega^2=d\theta^2+\sin^2\theta\, d\phi^2$. In these frames we have  $g^{00}=1$, $g_{0r}=g^{0r}=h$ and 
$ \sqrt{g}=\sqrt{-\det(g_{\mu\nu})}=r^2\sin\theta$ which may simplify some calculations. 

For the concrete dynamical charged systems with spherical symmetry, the functions $h$ have to be derived solving the Einstein-Maxwell equations with perfect fluid in the sense of selecting  the  manifolds ${\frak M}$ whose $h$-functions give  Einstein tensors that can be expressed as  
\begin{equation}\label{Ein}
	G^{\mu}_{\,\nu}=\Lambda\delta^{\mu}_{\nu}+ 8\pi T^{\mu}_{\nu}\,, \quad T^{\mu}_{\,\nu}={T_{fl}}^{\mu}_{\,\nu}+{T_{em}}^{\mu}_{\,\nu}\,,
\end{equation}
in terms of a possible cosmological constant $\Lambda$ and  the total stress-energy tensor, $T^{\mu}_{\,\nu}$,  formed by the  contribution of the perfect fluid, ${T_{fl}}^{\mu}_{\,\nu}$, and that of the electromagnetic field ${T_{em}}^{\mu}_{\,\nu}$. Assuming that te perfect fluid of density $\rho$ and pressure $p$ moves with the four-velocity $U^{\mu}$ with respect to the physical  frame under consideration we may write 
\begin{equation}\label{tfl}
	{T_{fl}}^{\mu}_{\,\nu}=	 (\rho+p)U^{\mu}U_{\nu}-p\delta^{\mu}_{\nu}\,.
\end{equation}
The contribution of the electromagnetic field,
\begin{equation}\label{Tem}
	{T_{em}}^{\mu}_{\,\nu}=\frac{1}{4\pi}\left(F^{\mu}_{\alpha}F^{\alpha}_{\nu}-\frac{1}{4}\delta^{\mu}_{\nu} F^{\beta}_{\alpha}F^{\alpha}_{\beta} \right)\,,
\end{equation}
depends on the field strength $F_{\mu\nu}=\partial_{\mu}A_{\nu}-\partial_{\nu}A_{\mu}$ derived from the electromagnetic potential $A_{\mu}$. 

For solving the Einstein-Maxwell equations it is convenient to consider the charged dynamical system in its proper co-moving frame where the four-velocity has the components 
\begin{eqnarray}
	U_{\mu}&=&\left(\frac{1}{\sqrt{g^{00}}},0,0,0\right)=(1,0,0,0)\,, \\ 
	U^{\mu}&=&g^{\mu 0}U_0=\frac{g^{\mu 0}}{\sqrt{g^{00}}} ~~\Rightarrow~~U^{\mu}=(1,h,0,0)\,,
\end{eqnarray}
assuming that in this frame the electromagnetic potential is produced by central electric charges, having the general form 
\begin{equation}\label{pot}
	(A_{\mu})=(A_0(t,r),0,0,0)\,,
\end{equation}    
defined up to a gauge.

The asymptotic behaviour in the proper co-moving frame defines the asymptotic space-time of  ${\frak M}$, for $r\to\infty$, which  is a FLRW manifold with flat space sections, ${\frak M}(a)$, of the scale factor $a(t)$, whose Hubble function is defined by the asymptotic condition 
\begin{equation}\label{Hub1}
	\frac{1}{a(t)}\frac{da(t)}{dt}\equiv\frac{\dot a(t)}{a(t)}=\lim_{r\to\infty}\frac{h(t,r)}{r}\,.
\end{equation}
The metric tensor of  ${\frak M}(a)$ gives the  line element in the physical proper frame, 
\begin{eqnarray}
	ds^2&=&g_{\mu\nu}(a)dx^{\mu}dx^{\nu}\nonumber\\
	&=&\left(1-\frac{\dot a^2}{a^2}\, r^2\right)dt^2+2\frac{\dot a}{a}\, r\, dr\, dt -dr^2-r^2d\Omega^2\,,\label{s2}
\end{eqnarray}
which emphasis an apparent horizon on the sphere of radius  
\begin{equation}\label{ra}
	r_a(t)=\left| \frac{a(t)}{\dot a(t)}\right| \,,
\end{equation}
we call the asymptotic horizon. Moreover, this satisfies the Friedmann equations with the isotropic Einstein tensor of components
\begin{eqnarray}
	G^0_{0}(a) &=&3\left(\frac{\dot a}{a}\right)^2=\Lambda + 8\pi \,\rho_a\,, \label{E1}\\
	G(a)&\equiv&G^r_r(a)=G^{\theta}_{\theta}(a)=G^{\phi}_{\phi}(a)\nonumber\\
	&=&3\left(\frac{\dot a}{a}\right)^2 +2\frac{d}{dt}\left(\frac{\dot a}{a}\right) =\Lambda - 8\pi \, p_a \,,\label{E2}
\end{eqnarray}
which satisfy the condition  
\begin{eqnarray}
	G^0_r(a)=0 ~~\Rightarrow~~G^r_0(a)=\frac{g^{0r}(a)}{g^{00}(a)}\left[G^0_0(a)-G(a)\right]\,.\label{cond}
\end{eqnarray}
Hereby, we deduce the asymptotic density $\rho_a$ and pressure $p_a$ that can be measured in the asymptotic zone of the proper frame, for  $r\to\infty$.

Furthermore, we ought to derive the stress-energy tensor in ${\frak M}$ but, unfortunately, this cannot be done because of some major technical difficulties. Under such circumstances, we adopt the same strategy as in the case of the Reissner-Nordstr\" om theory, restricting ourselves to evaluate the stress-energy tensor in the asymptotic space-time ${\frak M}(a)$ instead of ${\frak M}$. This can be derived as in Ref. \cite{CM} starting with the non-vanishing covariant components $F_{r,0}=-F_{0,r}=\partial_r A_0$ giving the components $F^{\mu}_{\,\nu}=g^{\mu\alpha}(a)F_{\alpha\nu}$ that help us to derive the definitive form of the tensor (\ref{Tem}) as
\begin{equation}
	\left({T_{em}}^{\mu}_{\nu}\right)=\frac{1}{8\pi}\Theta\,{\rm diag}(1,1,-1,-1)\,,
\end{equation} 
where
\begin{equation}
	\Theta(t,r)=\left[ \partial_rA_0(t,r)\right]^2\,,
\end{equation}
With these preparations, we may put the Einstein-Maxwell equations in the definitive form
\begin{eqnarray}
	G^0_0-\Theta&=& \Lambda+8\pi \rho\,,\label{EM1}\\
	G^r_r-\Theta&=&	G^{\theta}_{\theta}+\Theta=G^{\phi}_{\phi}+\Theta\nonumber\\
	&=&\Lambda-8\pi p\,,\label{EM2}\\
	G^0_r=0 &\Rightarrow&G^r_0=\frac{g^{0r}}{g^{00}}\left(G^0_0-G^r_r\right)\,,\label{EM3}
\end{eqnarray}
giving the expressions of the density $\rho$ and pressure $p$ of the perfect fluid of ${\frak M}$. Separating the parameters of the asymptotic FLRW fluid, 
\begin{equation}
	\rho(t,r)=\rho_a(t)+\delta\rho(t,r)\,, \quad p(t,r)=p_a(t)+\delta p(t,r)\,,	
\end{equation}
we obtain the image of a dynamical system (or particle) giving rise to the partial density $\delta \rho$ and pressure $\delta p$,  hosted by a FLRW fluid of parameters $(\rho_a,p_a)$.  

In the physical frames of the above geometries, the coordinate $t$ is the cosmic time only in the physical domains where  $g_{00}(t,r)>0$. This means that the physical space domain is delimited by two dynamical spherical horizons, the black hole and cosmological ones, whose radii, $r_b(t)$ and respectively $r_c(t)$,  have to be derived solving the equation $h(t,r)=1$ for expanding geometries and $h(t,r)=-1$ for collapsing ones \cite{Cot}. In many cases  these equations  have the desired positive solutions  only after a critical instant $t_{cr}$.  These horizons whose radii respect the hierarchy  $0<r_b(t)<r_c(t)<r_a(t)$ define a dynamical black hole  that can be observed only inside the physical domain. 

\section{Models of dynamical charged particles}

Let us look now for new solutions of the Einstein-Maxwell equations describing charged dynamical particles in their proper physical frames with metrics of the form  (\ref{fam}), having flat space sections.  Let us consider that the dynamical particle has the dynamical mass $M(t)$ and charge $Q(t)$. Then the Coulomb potential  gives \cite{CM}
\begin{equation}
	A_0(t,r)=\frac{Q(t)}{r}~~~\Rightarrow~~~\Theta=\frac{Q(t)^2}{r^4}\,,
\end{equation}
which suggests us to correct the $h$-functions of the $\kappa$-models with the quantity $A_0(t,r)^2$ as 
\begin{equation}\label{hany}
	h(t,r)=N(t) r+\epsilon\sqrt{\frac{2 M(t)}{r}+\Omega(t)^2 r^2-\frac{Q(t)^2}{r^2}}\,. 
\end{equation}
We remind the reader that the  function 
\begin{equation}
	\Omega(t)=|\kappa| M(t)\,,	\quad \kappa=\epsilon|\kappa|\,,
\end{equation}
simulates a Hubble-de Sitter dynamical parameters while  \cite{Cot1}
\begin{eqnarray}
	N(t)=-\frac{1}{3M(t)}\frac{dM(t)}{dt}=-\frac{1}{3}\frac{\dot{M}(t)}{M(t)}\,.\label{NM}
\end{eqnarray}
The $h$-functions (\ref{hany}) with arbitrary functions $Q(t)$ give  diagonal Einstein tensors but which, in general, cannot be put in the desired form  (\ref{EM1}-\ref{EM3}). Therefore, we must impose the supplemental condition 
\begin{equation}
	G^r_r-G^{\theta}_{\theta}=2\Theta=2 \frac{Q(t)^2}{r^4}\,,	
\end{equation}
which is accomplished if and only if we set
\begin{equation}
	Q(t)=\lambda M(t)^{\frac{1}{3}}\,,\label{Q}	
\end{equation}	
where $\lambda$ is the new real-valued parameter of these new solutions of the Einstein-Maxwell equations.

As these models are  determined by the mass functions $M(t)$  and the real-valued parameters $\kappa=\epsilon|\kappa|$ and $\lambda$  we say that these are $(\kappa,\lambda)$-models in space-times denoted  by ${\frak M}(M,\kappa,\lambda)$ understanding that these have $h$-functions of definitive form
\begin{eqnarray}\label{hdef}
	h_{(\kappa,\lambda)}(t,r)=-\frac{1}{3}\frac{\dot{M}(t)}{M(t)} r+\epsilon\sqrt{\frac{2 M(t)}{r}+\kappa^2 M(t)^2 r^2-\lambda^2 \frac{M(t)^{\frac{2}{3}}}{r^2}}\,.
\end{eqnarray}
In these space-times the  Coulomb potential is produced by the charge (\ref{Q}) such that
\begin{equation}
	A_0(t,r)=\lambda \frac{M(t)^{\frac{1}{3}}}{r} ~~\Rightarrow~~ \Theta\equiv\Theta(M,\lambda)=\lambda^2 \frac{M(t)^{\frac{2}{3}}}{r^4}\,.
\end{equation}
Under such circumstances, the Einstein tensors of the $(\kappa,\lambda)$-models have components $G^{\mu}_{\nu}(M,\kappa,\lambda)$ which satisfy the equations (\ref{EM1}) and (\ref{EM2}) that now read 
\begin{eqnarray}
	&&G^0_0(M,\kappa,\lambda)-\Theta(M,\lambda)=3\kappa^2 M(t)^2 +\frac{1}{3} \frac{\dot M(t)^2}{M(t)^2}-2\dot{M}(t)K(t,r)=8\pi \rho_{(\kappa,\lambda)}\,,\label{E1}\\
	&&G^r_r(M,\kappa,\lambda)-\Theta(M,\lambda)=G^{\theta}_{\theta}(M,\kappa,\lambda)+\Theta(M,\lambda)=G^{\phi}_{\phi}(M,\kappa,\lambda)+\Theta(M,\lambda)\nonumber\\
	&&\hspace*{26mm}=3\kappa^2 M(t)^2+ \frac{\dot M(t)^2}{M(t)^2}-\frac{2}{3} \frac{\ddot M(t)}{M(t)}=-8\pi p_{(\kappa,\lambda)}	\,,\label{E2}
\end{eqnarray}
where
\begin{equation}
	K(t,r)=\epsilon	\frac{3r(1+{M(t)}\kappa^2r^{3})-\lambda^2M(t)^{-\frac{1}{3}} }{3r^2\sqrt{M(t)^2\kappa^2r^4+2M(t)r -\lambda^2 M(t)^{\frac{2}{3}}}}\,.
\end{equation}
These equations define the density $\rho_{(\kappa,\lambda)}$ and the pressure $p_{(\kappa,\lambda)}$ of the  perfect fluids of the $(\kappa,\lambda)$-models.  Of course, the condition (\ref{EM3}) also is satisfied. 

As the Coulomb contribution in $h_{(\kappa,\lambda)}$  is negative we must verify if the natural physical condition 
\begin{equation}
	h_{(\kappa,\lambda)}(t,r)\in \mathbb{R} \Rightarrow	M(t)^2\kappa^2r^4+2M(t)r -\lambda^2 M(t)^{\frac{2}{3}}\ge 0
\end{equation}
is fulfilled. Solving the corresponding equation we observe that its greater positive solution represent  the minimal value, $r_{min}(t)$, for which this condition is still accomplished. Remarkably, if we assume that these solutions  have the form 
\begin{equation}\label{rmin}
	r_{min}(t)=\frac{c(\kappa,\lambda)}{M(t)^{\frac{1}{3}}}\,,
\end{equation}
we find that the coefficients $c(\kappa,\lambda)$ satisfy the  simpler static equation
\begin{equation}\label{Emin}
	c(\kappa,\lambda)^4	\kappa^2+2c(\kappa,\lambda)-\lambda^2=0\,.
\end{equation}
We obtain thus the domain
\begin{eqnarray}
	{\frak D}_t=\{t,r, \theta, \phi\, |\,r\ge r_{min}(t)\} \,,
\end{eqnarray}  
in which the function $h_{(\kappa,\lambda)}(t,r)$ takes real values. In what follows we say that the sphere of radius $r_{min}$ is the {\em prohibited} sphere bearing in mind that this must be excluded from the space-time domain because in its interior the metric takes complex values.

The asymptotic space-time of ${\frak M}(M,\kappa,\lambda)$ is the FLRW space-time ${\frak M}(a)$ whose Hubble function  defined by Eq. (\ref{Hub1}) reds now 
\begin{equation}\label{at}
	\frac{\dot a(t)}{a(t)}=\lim_{r\to\infty}\frac{ h_{(\kappa,\lambda)}(t,r)}{r}=\kappa M(t) -\frac{1}{3}\frac{\dot{M}(t)}{M(t)}\,.
\end{equation}
Integrating this equation  with the initial condition $a(t_0) =1$ we obtain the scale factor  
\begin{equation}\label{at1}
	a(t)=\left(  \frac{M_0}{M(t)}\right)^{\frac{1}{3}} \exp\left( \kappa\int_{t_0}^t M(t')dt'\right)
\end{equation}
of the asymptotic FLRW space-time  where we denoted $M_0=M(t_0)$.

The Hubble functions must depend monotonously of time, without zeros in the physical time domain which  might produce singularities of the function $r_a(t)$ giving the radius of the asymptotic horizon (\ref{ra}).  For avoiding these zeros we  impose specific conditions for expanding or collapsing  space-times as 
\begin{eqnarray}
	{\rm expanding:} &~~~~&\frac{\dot a(t)}{a(t)}>0 ~~\Rightarrow ~~ 	\frac{\dot{M}(t)}{M(t)}<0\,,\quad \kappa>0\,,\label{expand}\\
	{\rm collapsing:} &~~~~&	\frac{\dot a(t)}{a(t)}<0 ~~\Rightarrow ~~ 	\frac{\dot{M}(t)}{M(t)}>0\,,\quad \kappa<0\,.
\end{eqnarray} 
In what follows we restrict ourselves to the expanding space-times which are of interest in cosmology. 

The physical meaning of the $(\kappa$, $\lambda)$-models results from the right-handed sides of the Einstein-Maxwell equations (\ref{E1}) and (\ref{E2}). First of all, we observe that there are no static terms which means that we do not need to consider a cosmological constant, setting $\Lambda=0$. Furthermore, we separate the density and pressures of the FLRW fluid 
\begin{eqnarray}
	\rho_a(t)&=&\frac{3}{8\pi}\left(\frac{\dot a}{a}\right)^2  \nonumber\\
	&=&\frac{1}{8\pi}\left( 3\kappa^2 M(t)^2 +\frac{1}{3} \frac{\dot M(t)^2}{M(t)^2}-2\kappa \dot M(t)\right)\,,\\
	p_a(t)&=&- \frac{1}{8\pi}\left[ 3\left(\frac{\dot a}{a}\right)^2 +2\frac{d}{dt}\left(\frac{\dot a}{a}\right)   \right]  \nonumber\\
	&=&-\frac{1}{8\pi}\left( 3\kappa^2 M(t)^2+ \frac{\dot M(t)^2}{M(t)^2}-\frac{2}{3} \frac{\ddot M(t)}{M(t)} \right)\,,
\end{eqnarray}
which depend only on the Hubble function (\ref{at}). Hereby it results that the total density $\rho_{(\kappa,\lambda)}=\rho_a + \delta\rho$  of the perfect fluid of the space-time ${\frak M}(M,\kappa,\lambda)$ gets the new point-dependent  term 
\begin{equation}
	\delta\rho(t,r)=\frac{1}{4\pi} \dot M(t)\left[  \kappa  - K(t,r)  \right] \,, \label{dust}	
\end{equation}
while its pressure remains unchanged, $p_{(\kappa,\lambda)}=p_a$. This means that $\delta\rho$ is the density of an amount of {\em dust} which does not modify te pressure $p_a$ of the FLRW fluid.  The total dust mass \cite{Cot2}
\begin{eqnarray}
	\delta M(t)=4\pi \int_{0}^{\infty} dr\,r^2 \delta\rho(t,r) =\left\{ 
	\begin{array}{cl}
		-\frac{1}{3\kappa}\frac{\dot M(t)}{M(t)}& ~~~\lambda=0\,,\\	
		\infty&~~~~\lambda\not=0\,,
	\end{array}
	\right.
\end{eqnarray}
is finite only for $\lambda=0$ and $\kappa\not=0$.  Otherwise this is infinite just as the total mass of the FLRW perfect fluid.

We obtained thus a system formed by a dynamical particle (or black hole) of Reissner-Nordstr\" om type, surrounded by a cloud of dust hosted by the homogeneous perfect fluid of a FLRW space-time. It is remarkable that the principal features of all these components are determined only by the function $M(t)$ and the parameters $\kappa$ and $\lambda$. The black hole is produced by the typical  singular term  $\frac{2M(t)}{r}-\frac{Q(t)^2}{r^2}$ of the function (\ref{hdef}) while the dust density (\ref{dust}) was obtained solving the Einstein-Maxwell equations.

In general, these dynamical particles are black holes having  black hole and cosmological horizons whose radii, $r_b(t)$ and respectively $r_c(t)$  are positive solutions of the equation $h(t,r)=1$ that can be written as $[h(t,r)-rN(t)]^2 -[1-rN(t)]^2=0$ taking the final form
\begin{eqnarray}
	\left[ \kappa^2 M(t)^2 -N(t)^2\right]r^4+2N(t)r^3
	-r^2+2 M(t)r-\lambda^2	M(t)^{\frac{2}{3}}=0\,.\label{horE}
\end{eqnarray}
This equation has two physical solutions representing the horizon radii and two non-physical ones. As mentioned before the horizons arise on the same sphere at the critical time $t_{cr}$ when $r_b(t_{cr})=r_c(t_{cr})$ evolving then in opposite directions, $r_c(t)$ increasing to $r_a(t)$ and $r_b(t)$ collapsing to zero. We obtain thus the domain 
\begin{eqnarray}
	{\frak H}_t=\{t,r, \theta, \phi\, |\,t>t_{cr}, r_b(t)<r<r_c(t)\} \,,
\end{eqnarray}  
between the horizons of  ${\frak M}(M,\kappa,\lambda)$ where $t$ is time-like.   Bering in mind that, in addition, we must have $r>r_{min}(t)$ we understand that the physical domain where an observer can measure the black hole evolution is
\begin{equation}
	{\frak P}_t=	{\frak H}_t \cap	{\frak D}_t\,.	
\end{equation}
If  $r_{min}(t)<r_b(t)$   the function $h_{(\kappa,\lambda)}$ takes real values on the whole  domain $	{\frak H}_t$, the prohibited  sphere  remaining behind the black hole horizon. Otherwise, if $r_{min}(t)>r_b(t)$, then the prohibited sphere  takes over the role of the black hole horizon and the physical domain gets a new dynamics as we shall see in the following examples.  

\section{Examples}

The simplest examples are the static solutions with $M(t)=m=$ const. describing usual Reissner-Nordstr\" om-de Sitter geometries with the metric given by functions of the form (\ref{frSdS}) with $\omega_H=\kappa m$ and $q=\lambda m^{\frac{1}{3}}$.

The dynamical models with $\kappa=\lambda=0$ are Schwarzschild-type dynamical black holes \cite{Cot} while for $\lambda=0$ and $\kappa\not=0$ we obtain the $\kappa$-models describing neutral dynamical black holes we studied recently \cite{Cot1,Cot2}. The new models we intend to inspect here are those with $\lambda\not=0$,  describing charged dynamical particles. Our principal objective is to verify that these particles are charged black holes that can be observed in the physical domains bordered by their inner and outer  horizons. As here we study for the first time these models we try to avoid the numerical methods we need for solving Eq. (\ref{horE}) with arbitrary parameters in the favor of analytical ones that can be applied when this equation becomes a cubic one after dropping out its first term.

Therefore, in what follows we focus on the models of expanding geometries whose parameters comply with the condition 
\begin{equation}\label{cMt}
	\kappa^2 M(t)^2 -N(t)^2=0 ~~ \Rightarrow~~	 \kappa^2 M(t)-\left(\frac{1}{3}\frac{\dot M(t)}{M(t)}\right)^2=0\,.
\end{equation}
Solving these equations we find only one physical solution, 
\begin{equation}
	M(t)=\frac{M_0}{1+3\kappa M_0 (t-t_0)}\,,
\end{equation} 
which satisfies the initial condition $M(t_0)=M_0>0$. In addition, for avoiding unwanted singularities, we assume that these models have $\kappa>0$ and arbitrary $\lambda \in\mathbb{R}$.

{\begin{figure}
		\centering
		\includegraphics[scale=0.35]{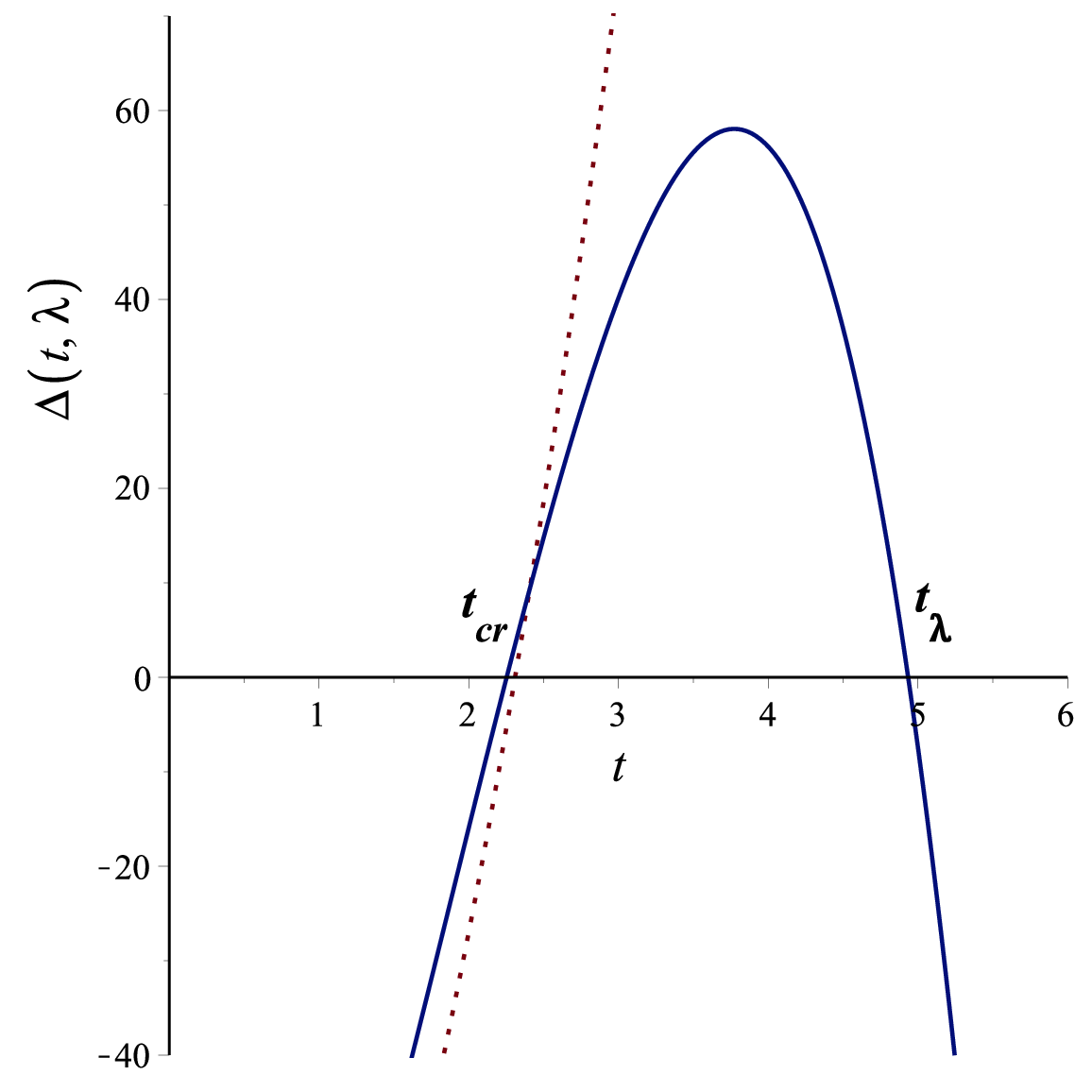}
		\includegraphics[scale=0.35]{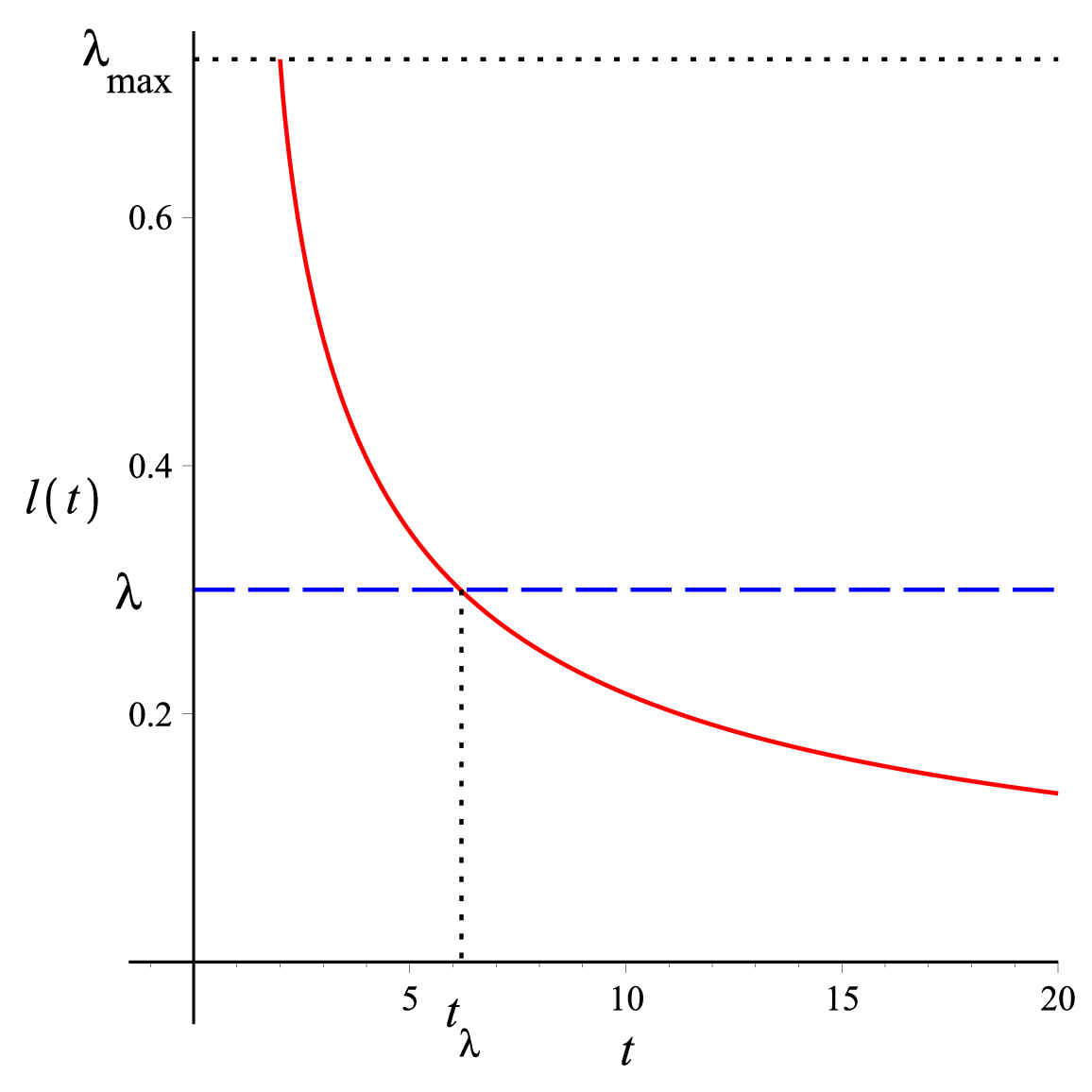}	
		\caption{In the left panel we plot the function $\Delta(t, 0)$ (dotted line) and $\Delta(t, 0.35)$ (solid line) which has a second root for $t=t_{\lambda}$.  In the right panel we see the function $l(t)$ and how $t_{\lambda}$ can be derived graphically. }
\end{figure}} 

For a brief graphical analysis it is convenient to simplify the parametrization setting
\begin{equation}
	t_0=1\,,~~~~ M_0=1\,, ~~~~ \kappa=\frac{1}{3}~~ \Rightarrow ~~ M(t)=\frac{1}{t}\,, 	
\end{equation}
obtaining a mass function with a singularity at $t=0$ which can be interpreted as the big bang time.  Remarkably, all the models with these parameters and arbitrary $\lambda \in\mathbb{R}$  have the same asymptotic manifold ${\frak M}(a)$ which is just the matter-dominant universe with
\begin{equation}
	a(t)=t^{\frac{2}{3}}\,, ~~~\Rightarrow~~~ r_a(t)=\frac{3}{2}\, t\,,~~~~t\in [0,\infty)\,,
\end{equation}
as it results from  Eq. (\ref{at1}). Eq. (\ref{horE}) giving the horizon radii  becomes now the cubic equation
\begin{equation}\label{cE}
	2r(t)^3-3 tr(t)^2+6 r(t)-3\lambda^2t^{\frac{1}{3}}=0\,,
\end{equation}
that can be analytically solved, allowing  three real-valued solutions when its discriminant 
\begin{equation}\label{fD}
	\Delta(t,\lambda)=\frac{27}{4}\left[3t^2-9t^{\frac{2}{3}}\lambda^4+t^{\frac{4}{3}}\left(18-3t^2\right)\lambda^2-16\right]\,,
\end{equation}
is positively defined and only one when this is negative.   In the first case only two solutions have physical meaning, the horizon radii $r_b(t)$ and $r_c(t)$, while the third solution, $r_{np}(t)$,  does not have a physical interpretation. When the discriminant becomes negative we remain only with  the radius of the cosmological horizon, $r_c(t)$. 

{ \begin{figure}
		\centering
		\includegraphics[scale=0.35]{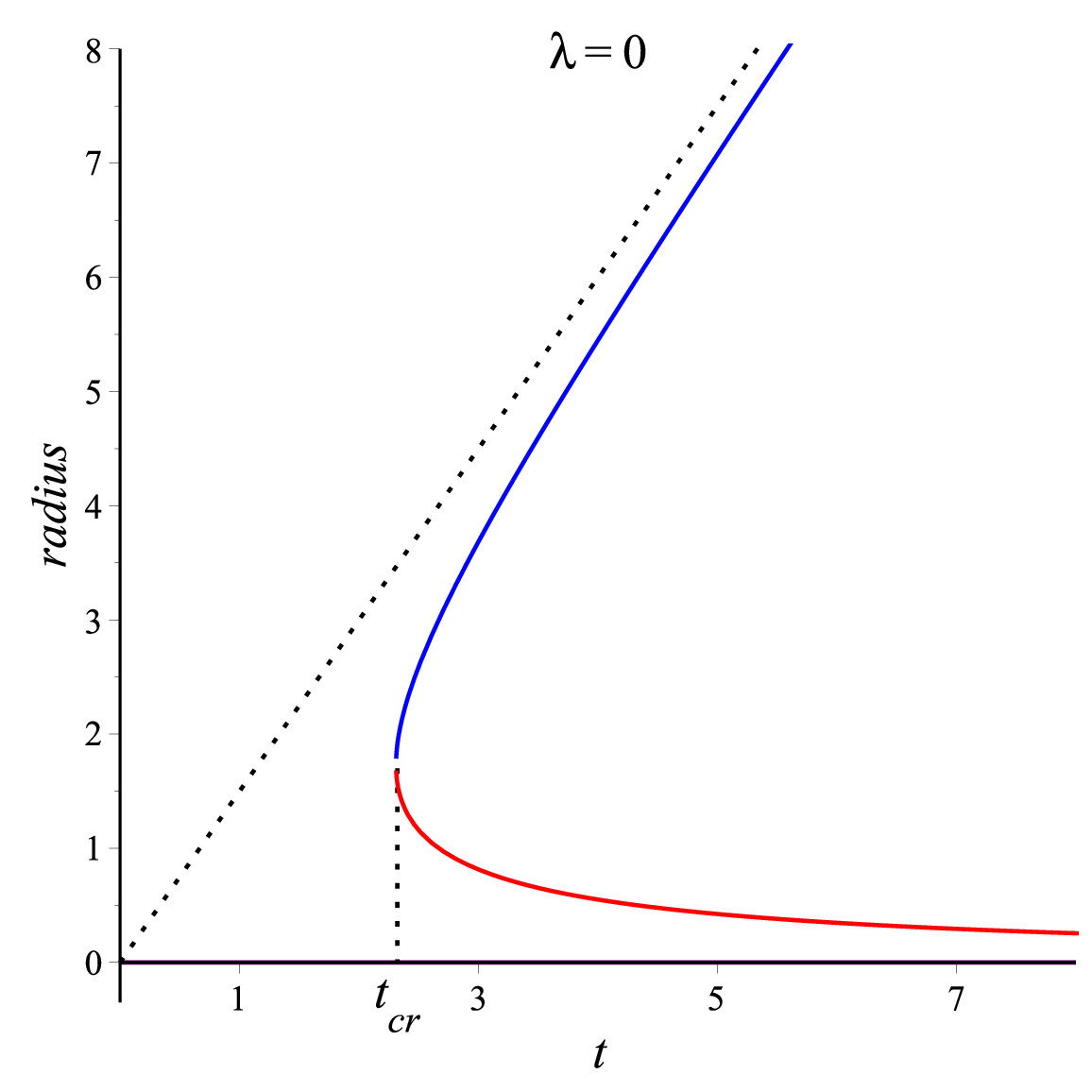}
		\includegraphics[scale=0.35]{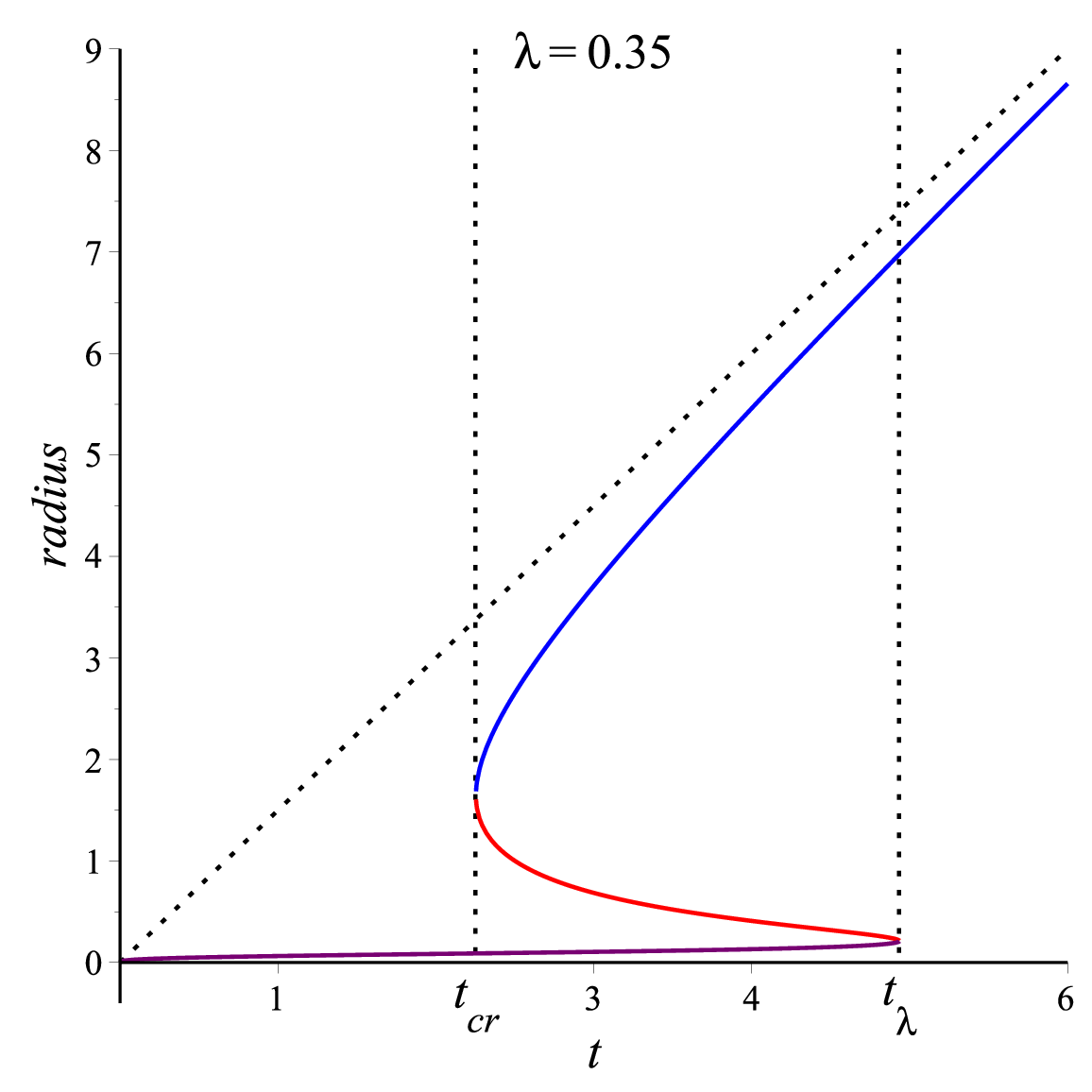}	
		\caption{The functions $r_c(t)$ and $r_b(t)$ forming a complete C-curve for $\lambda=0$ (left panel) or an incomplete one for $\lambda=0.35$ but which is connected with the non-physical solution $r_{np}(t)$ taking  real values for $0<t<t_{\lambda}$  (right panel).}
\end{figure}}

Therefore it is important to find the time domain in which the discriminant is positive. For all the models we discuss here, the discriminant becomes positive at the critical time $t_{cr}$ when the black hole and cosmological horizons arise on the same sphere of radius $r_b(t_{cr})=r_c(t_{cr})$. In the case of $\lambda=0$ this remains positive when $t$ increases to infinity but if $\lambda\not= 0$ it change the sign again at the time $t_{\lambda}$ whose value depends on $\lambda$ (see the left panel of Fig. 1). This instant can be derived either solving directly the equation $\Delta(t_{\lambda},\lambda)=0$ or exploiting the auxiliary function 
\begin{equation}\label{lt}
	l(t)=\left[\frac{1}{6 t^{\frac{1}{3}}} \left(-t^3+6 t+\sqrt{(t-2)^3(t+2)^3}\right)   \right]^{\frac{1}{2}}\,,
\end{equation}
that is the positive solution of the equation  $\Delta(t,l(t))=0$  (which has a pair of real-valued solutions, $\pm l(t)$, and a pair of complex conjugated ones). We observe that this function is defined on the domain $t\in[2,\infty)$ decreasing monotonously to zero (as in the right panel of Fig. 1). Its maximal value $\lambda_{max}=l(2)=  0.727415$ gives the interval $[-\lambda_{max}, \lambda_{\max}]$ in which the parameter $\lambda$ has physical meaning. This means that a charged black hole of mass $M(t)$ may carry at most the maximal charge  $|Q_{\max}(t)|=\lambda_{max}M(t)^{\frac{1}{3}}$. On the other hand, the function (\ref{lt}) helps us to determine the time $t_{\lambda}$ when the discriminant of a model with given $\lambda$ changes its sign solving the equation $l(t_{\lambda})=|\lambda|$ as in the right panel of Fig. 1. We  identify thus  three time domains,
\begin{eqnarray}
	t<t_{cr}&\Rightarrow&\Delta(t,\lambda)	<0\nonumber\\
	&\Rightarrow& r_{b}(t),\, r_c(t)\in \mathbb{C},~~~~ r_{np}(t) \in \mathbb{R}\,,\nonumber\\	
	t_{cr}<t<t_{\lambda} &\Rightarrow&\Delta(t,\lambda)	>0\nonumber\\
	&\Rightarrow& r_{np}(t)<r_b(t)<r_c(t) \in \mathbb{R}\,,\label{123}\\
	t>t_{\lambda} &\Rightarrow&\Delta(t,\lambda)	<0\nonumber\\
	&\Rightarrow&r_{np}(t),\, r_b(t)\in \mathbb{C},~~~~ r_c(t) \in \mathbb{R}\,,\nonumber
\end{eqnarray}
where the horizons form an incomplete C-curve as in the right panel of Fig 2.

{ \begin{figure}
		\centering
		\includegraphics[scale=0.35]{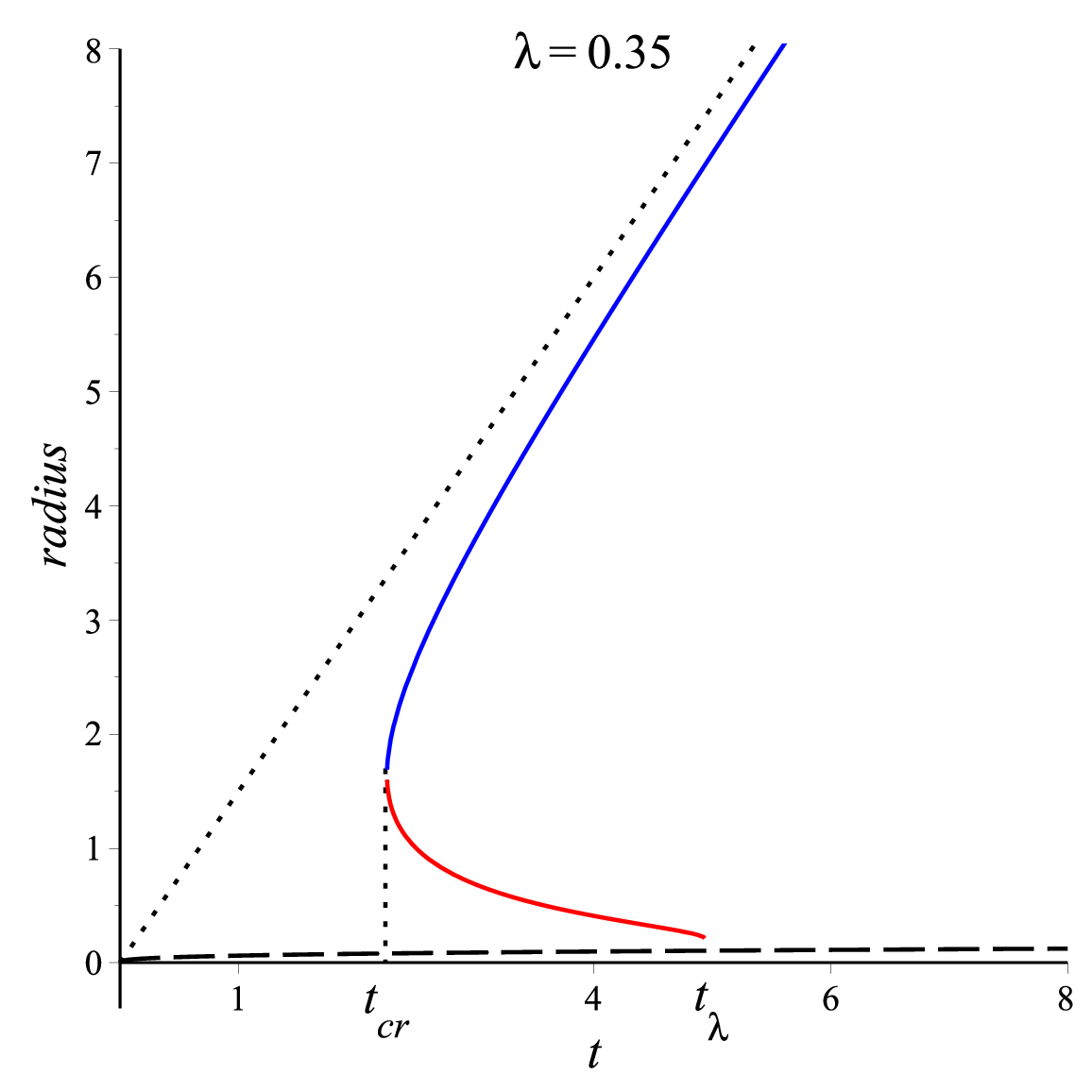}
		\includegraphics[scale=0.35]{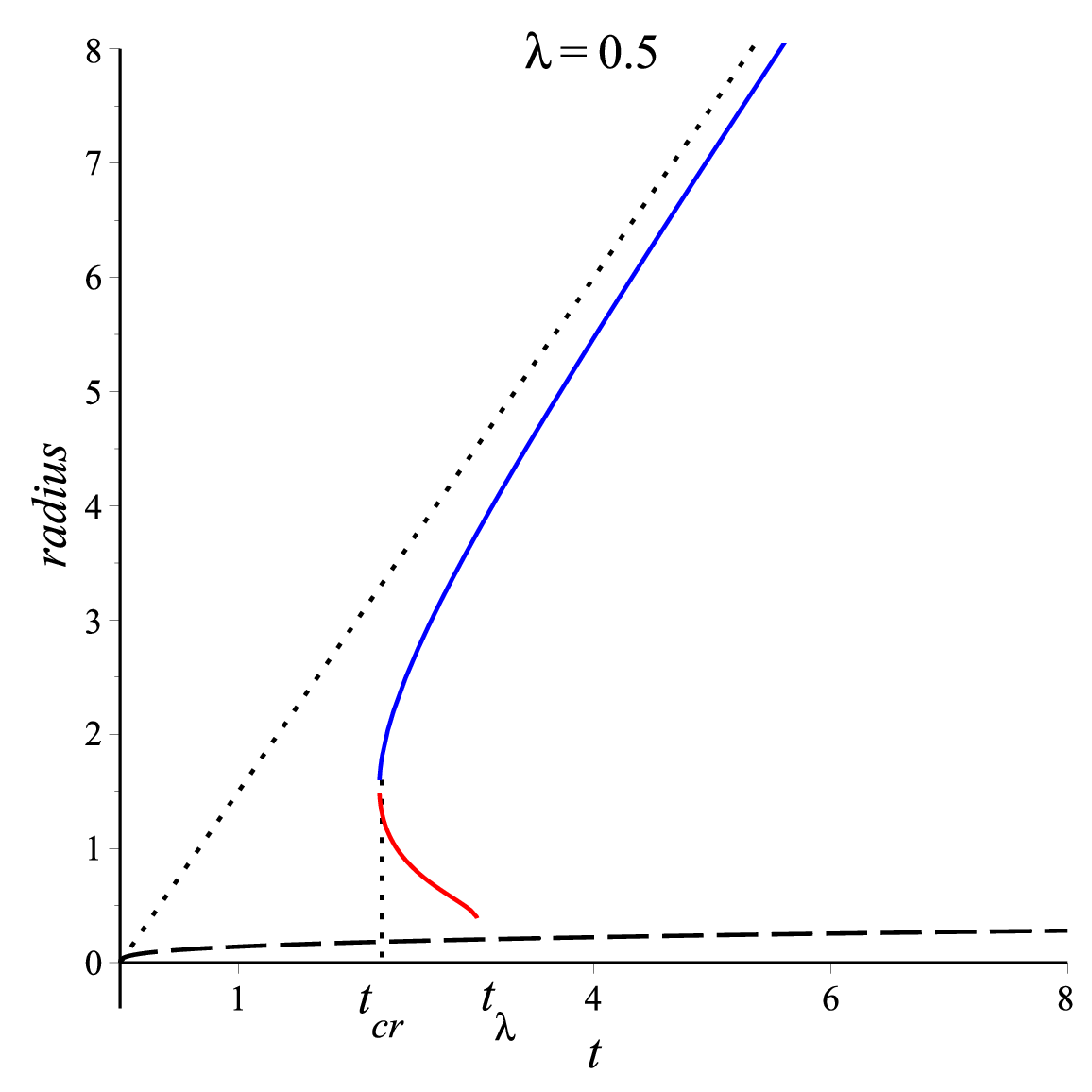}	
		\caption{The functions $r_{max\,1}$ (left panel) and $r_{max\,2}$ (right panel), plotted with dashed lines, complete the borders of the physical domains of the models with $\lambda\not= 0$.}		
\end{figure}}

For $t>t_{\lambda}$ we remain with the restriction $r>r_{min}(t)$ which means that the prohibited sphere  takes over the role of the missing black hole horizon. In the case of  our models, the function  $r_{min}(t)$ has the general form (\ref{rmin}) with the constant $c(\kappa,\lambda)$ that may be derived numerically as the greater real solution of  Eq. (\ref{Emin}).  Here we restrict ourselves to the following examples
\begin{eqnarray}
	\kappa={\textstyle \frac{1}{3}} \,, ~~\lambda=0.35 &\Rightarrow& c_1\equiv c({\textstyle \frac{1}{3}},0,35)=0.061225\,,\\
	\kappa={\textstyle \frac{1}{3}} \,, ~~\lambda=0.5~~ &\Rightarrow& c_2\equiv c({\textstyle \frac{1}{3}},0,5)~~=0.139969\,,	
\end{eqnarray}
giving the functions $r_{min\, 1}(t)=c_1 t^{\frac{1}{3}}$ and $r_{min\, 2}(t)=c_2  t^{\frac{1}{3}}$ that complete the profiles of the physical domains in both these cases  of incomplete C-curves. Note that these functions are increasing monotonously in time in contrast with the radius of the black hole horizon of the model with $\lambda=0$ which collapses to zero as in the left panel of the Fig. 2. 

We hope that these analytical results may be a guide in further investigations of models for which Eq. (\ref{horE}) is cuartic and we are forced to use numerical methods.

\section{Concluding remarks}

We conclude  that our new $(\kappa, \lambda)$-models  describe systems formed by a Reissner-Nordstr\" om-type dynamical charged black hole  surrounded by a cloud of dust hosted by the perfect fluid of the asymptotic FLRW space-time. Any such system gets physical meaning  at a critical instant $t_{cr}$ when the bkack hole and cosmological  horizons appear on the same sphere,  evolving then in opposite directions, creating the physical space between their spheres for $t<t_{\lambda}$. At the instant $t=t_{\lambda}$ the black hole horizon is replaced by the prohibited sphere of radius $r_{min}(t)$ which increases in time but without approaching to the cosmological horizon. 

The specific features of these models are somewhat unusual. The most  intriguing is the proportionality between the dynamical charge and the cubic root of the mass function whose physical meaning is not yet obvious. Another unusual property is the limitation of the dynamical charge carried by a black hole that seems to depend on the concrete model under consideration. For understanding the physical meaning of these new features  various models must be analyzed using  numerical methods for the models with $\kappa\not=\frac{1}{3}$. 

Moreover, it is crucial to study the quantities that can be measured by an observer in the physical domain of the black hole. In our opinion the principal problem we must consider with priority is that of the photon sphere for $t>t_{\lambda}$ when the prohibited sphere, which replaces the black hole horizon, is expanding if the whole space-time is expanding. Solving this problem we may obtain information about the time evolution of the black hole shadow that we know that may be related to the redshift \cite{Cbs}. 

Finally, we note that  this paper closes the series of studies of dynamical non-rotating black holes in asymptotic FLRW space-times \cite{Cot,Cot1,Cot2} that can be derived solving the Einstein or Einstein-Maxwell equations using $h$-functions of the general form (\ref{hany}).  However, we do not exclude the existence of other types of dynamical black holes that could be found in further investigations.

\appendix
\section{Solving cubic equations}

\setcounter{equation}{0} \renewcommand{\theequation}
{A.\arabic{equation}}

For solving the cubic equation 
\begin{equation}
	ar^3+br^2+cr+d=0\,, \label{e1}	
\end{equation}
we substitute first  $r=x-\frac{b}{3a}$ obtaining the modified depressed equation $x^3-p x+q=0$ with the coefficients
\begin{eqnarray}
	p=\frac{b^2-3ac}{3a^2}\,,\quad q=\frac{27a^2d-9abc+2b^3}{27a^3}\,,\label{pq}
\end{eqnarray}
that can be solved using the following form of  Cardano's formulae 
\begin{eqnarray}
	x_1&=&\frac{1}{2}\left(A+\frac{4}{3}\frac{p}{A}\right)\,,\\
	x_2&=&\frac{1}{2}\left(Ae^{i\frac{2\pi}{3}}+\frac{4}{3}\frac{p}{A}e^{-i\frac{2\pi}{3}}\right)\,,\\
	x_3&=&\frac{1}{2}\left(Ae^{i\frac{4\pi}{3}}+\frac{4}{3}\frac{p}{A}e^{-i\frac{4\pi}{3}}\right)\,,	
\end{eqnarray}
where $	A=\left(\frac{2}{3}\right)^{\frac{2}{3}}\left[ i\sqrt{3}\sqrt{4p^3-27 q^2}-9 q \right]^{\frac{1}{3}}$. The cubic equations allow real solutions  only when their discriminants are positive, $\Delta=4 p^3-27 q^2 >0$. 

For the models with $\kappa=\frac{1}{3}$  the equation (\ref{horE}) becomes a cubic one  (\ref{cE}) that has  the canonical form (\ref{e1}) with the coefficients 
\begin{eqnarray}	
	a(t)=2 t^{\frac{2}{3}}\,,~~~ b(t)=-3t^{\frac{5}{3}}\,,~~~
	c =6 t^{\frac{2}{3}}\,, ~~~ d(t)=-3\lambda^2 t\,,
\end{eqnarray}
giving the quantities (\ref{pq}) and the discriminant (\ref{fD}). Finally, the solutions we are looking for can be identified as
\begin{eqnarray}\label{rbc}
	r_c(t)&=&x_1(t)-\frac{b(t)}{3a(t)}\,,\\
	r_{np}(t)&=&x_2(t)-\frac{b(t)}{3a(t)}\,,\\ 
	r_b(t)&=&x_3(t)-\frac{b(t)}{3a(t)}\,.
\end{eqnarray}
These functions take real values as in Eqs. (\ref{123}).

\end{document}